%% file: paper.tex
\documentclass{article}

\usepackage{arxiv}

\usepackage[utf8]{inputenc} 
\usepackage[T1]{fontenc}    
\usepackage{hyperref}       
\usepackage{url}            
\usepackage{booktabs}       
\usepackage{amsfonts}       
\usepackage{nicefrac}       
\usepackage{microtype}      
\usepackage{cleveref}       
\usepackage{lipsum}         
\usepackage{graphicx}
\usepackage{natbib}
\usepackage{doi}
\usepackage{xspace}

\newcommand{\code}[1]{\mbox{\texttt{#1}}}
\newcommand{\toolregistry}{\mbox{ToolRegistry}\xspace}

\title{\toolregistry: A Protocol-Agnostic Tool Management Library for Function-Calling LLMs}

\newif\ifuniqueAffiliation
\uniqueAffiliationfalse

\ifuniqueAffiliation
\author{
\href{https://orcid.org/0000-0001-8353-0821}{\includegraphics[scale=0.06]{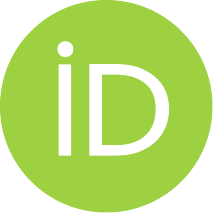}\hspace{1mm}Peng Ding} \\
University of Chicago \\
\texttt{dingpeng@uchicago.edu}
}
\else
\author{
\href{https://orcid.org/0000-0001-8353-0821}{\includegraphics[scale=0.06]{orcid.pdf}\hspace{1mm}Peng Ding} \\
University of Chicago \\
\texttt{dingpeng@uchicago.edu} \\
\And
Rick Stevens \\
University of Chicago \\
Argonne National Laboratory \\
\texttt{stevens@cs.uchicago.edu} \\
}
\fi

\renewcommand{\shorttitle}{ToolRegistry}

\hypersetup{
pdftitle={ToolRegistry: A Protocol-Agnostic Tool Management Library for Function-Calling LLMs},
pdfsubject={Computer Science - Software Engineering, Computer Science - Artificial Intelligence},
pdfauthor={Peng Ding, Rick Stevens},
pdfkeywords={LLM tools, function calling, protocol agnostic, tool registry, concurrent execution, permission system, tool discovery, think-augmented, MCP},
}

\begin{document}
\maketitle

\begin{abstract}
	Every LLM tool call is structurally an RPC---a function name, JSON arguments, and a serialized result---yet each protocol (native Python, MCP, OpenAPI, LangChain) is integrated from scratch. We present \toolregistry, a system that makes this RPC nature explicit: a single \code{Tool} object acts as a universal stub regardless of transport, while the registry serves as the RPC client runtime---handling dispatch, schema generation, concurrency, and error recovery. The system ships as three packages---a core registry, a server exposing tools over MCP and OpenAPI, and a hub of production-ready implementations---and invokes tools through pluggable thread or process backends. The system now also provides tag-based permission policies, BM25F-powered progressive tool disclosure for large registries, think-augmented function calling, multi-provider schema support (OpenAI, Anthropic, Gemini), declarative JSONC/YAML configuration, and a near-zero-dependency core built on stdlib-only vendored modules. In our benchmarks the library cuts integration code by 60--80\%, and choosing the right concurrency mode (thread vs.\ process) yields up to 3.1$\times$ throughput over the alternative for a given workload. \toolregistry is open-source at \url{https://github.com/Oaklight/ToolRegistry}; documentation lives at \url{https://toolregistry.readthedocs.io/}.
\end{abstract}

\keywords{LLM Tools \and Function Calling \and Tool Integration \and Protocol Agnostic \and Concurrent Execution \and Permission System \and Tool Discovery \and Think-Augmented \and MCP}

\input{sections/intro}

\input{sections/related_work}

\input{sections/design}

\input{sections/cases}

\input{sections/evaluation}

\input{sections/limitations}

\input{sections/conclusion}

\section*{Acknowledgments}

Large language models were used to assist with proofreading and language editing. The authors take full responsibility for all content.

\bibliographystyle{unsrtnat}
\bibliography{references}  






\end{document}

%% file: sections/intro.tex
\section{Introduction}

Large Language Model (LLM) applications increasingly delegate work to external tools---for data retrieval, computation, and interaction with real-world services. Structurally, every such invocation follows the Remote Procedure Call (RPC) pattern: the LLM emits a function name and JSON arguments, a runtime dispatches the call, and a serialized result is returned. Whether the callee is a local Python function (an in-process call), an MCP server (JSON-RPC over stdio or HTTP), an OpenAPI endpoint (REST over HTTP), or a LangChain tool (an in-process adapter), the calling convention is the same. Yet the ecosystem treats each transport as a separate integration problem, forcing developers to write protocol-specific glue for what is fundamentally the same operation.

This fragmentation exposes four interconnected challenges:
\begin{enumerate}
	\item \textbf{Protocol Fragmentation:} OpenAPI is stable and mature; MCP (Model Context Protocol) is gaining traction; many teams still use bare Python functions. No single standard covers all cases, so developers juggle multiple protocols.
	\item \textbf{Manual Schema Overhead:} Most frameworks require hand-written JSON schemas for every tool---parameter types, descriptions, constraints---even for trivial functions. The boilerplate dwarfs the actual logic.
	\item \textbf{Complex Execution Workflow:} Each framework has its own calling convention, message format, and sync/async mix. Parallelizing calls across these conventions demands non-trivial concurrency plumbing.
	\item \textbf{Provider Lock-In:} OpenAI's Chat Completion API is the de facto standard, but Anthropic and Gemini use different tool-call structures. Applications that target multiple providers maintain separate code paths for what is semantically the same operation.
\end{enumerate}

Starting from the observation that tool calling \emph{is} RPC, we introduce \toolregistry, a system that provides a universal stub---the \code{Tool} object---abstracting over transport and binding so that registration, schema generation, invocation, and lifecycle management work identically regardless of where or how a tool runs. Since its initial release as a single library, \toolregistry has evolved into a modular three-package ecosystem:
\begin{itemize}
	\item \textbf{toolregistry} (core, v0.10.1): Tool model, registry, schema generation, invocation engine, permission system, progressive tool discovery, and multi-provider API compatibility layer.
	\item \textbf{toolregistry-server} (v0.3.1): Protocol adapters (MCP, OpenAPI), route table, CLI, declarative configuration, and authentication.
	\item \textbf{toolregistry-hub} (v0.8.1): Curated, production-tested tool implementations (calculator, web search, datetime, filesystem, bash, and more) with JSONC configuration and Docker deployment support.
\end{itemize}

Unlike frameworks that impose their own architecture, \toolregistry acts as a helper library: it drops into existing codebases, wraps tools from any source in a single \code{Tool} object, and lets developers focus on application logic instead of integration boilerplate.

The main contributions are:
\begin{enumerate}
	\item Protocol-agnostic registration unifying native Python, MCP, OpenAPI, and LangChain tools under one interface
	\item Automated JSON-schema generation from type hints and docstrings, eliminating hand-written schemas
	\item Pluggable dual-mode invocation engine (thread pool for CPU-bound, process pool for I/O-bound workloads)
	\item Three-package ecosystem (core, server, hub) with event-driven change propagation and a near-zero-dependency core via stdlib-only vendored modules
	\item Web-based admin panel (bilingual EN/ZH) with execution logging, state export/import, and runtime metadata control
	\item Independent MCP client supporting four transports (stdio, SSE, streamable HTTP, WebSocket) with persistent connections
	\item Tag-based permission system: \code{ToolTag} classification, first-match-wins \code{PermissionPolicy}, five built-in rules
	\item Progressive tool disclosure via BM25F-based \code{ToolDiscoveryTool} with deferred loading for large registries
	\item Think-augmented function calling: injects a reasoning property into tool schemas, controllable per-tool or registry-wide
	\item Multi-provider API support (OpenAI, Anthropic, Gemini) plus declarative JSONC/YAML tool configuration
	\item Evaluation showing 60--80\% code reduction and up to 4.5$\times$ throughput difference between execution modes, guiding workload-appropriate concurrency selection
\end{enumerate}

The paper proceeds as follows: Section~2 surveys related work. Section~3 details system design---core abstractions, execution engine, permission system, progressive disclosure, think-augmented calling, and the ecosystem structure. Section~4 presents case studies. Section~5 reports performance and developer-experience metrics. Section~6 discusses limitations and future work, and Section~7 concludes.

%% file: sections/related_work.tex
\section{Related Work}

\subsection{Evolution of Tool-Augmented LLMs}

\citet{schick2023toolformer} showed that LLMs can learn tool usage through self-supervision. Follow-up work scaled the idea: \citet{qin2023toolllm} automated training over 16,000 real-world APIs, while \citet{lu2023chameleon} and \citet{shen2023hugginggpt} explored compositional reasoning and multi-tool orchestration. As tool pools grow, \emph{tool retrieval} becomes a bottleneck: \citet{du2024anytool} introduced hierarchical retrieval over large API sets, and \citet{shi2025toolret} benchmarked retrieval models on 43k tools, finding that even strong IR models perform poorly on tool selection---motivating the BM25F-based progressive disclosure mechanism in \toolregistry.

\subsection{Current Paradigms in Tool Learning}

Surveys \citep{shen2024llm, qu2025tool} group approaches into fine-tuning (Toolformer, Gorilla~\citep{patil2024gorilla}), in-context learning (Chameleon~\citep{lu2023chameleon}), and orchestration (HuggingGPT~\citep{shen2023hugginggpt}). In practice these converge: even fine-tuned models use standardized function-calling interfaces at inference time, making unified tool management both feasible and necessary.

\subsection{Protocol Standardization Challenges}

\subsubsection{Function Calling Standards}
OpenAI, Anthropic, and Google all expose JSON-based function calling, but their schema structures differ enough to require provider-specific code paths for schema generation, tool-call parsing, and result formatting. Projects like llm-rosetta~\citep{llmrosetta2025} tackle multi-provider translation as a standalone concern.

\subsubsection{Model Context Protocol (MCP)}
Anthropic's Model Context Protocol (MCP)~\citep{anthropic2024mcp} separates invocation logic from tool implementations. Its latest revision replaces SSE with Streamable HTTP~\citep{modelcontextprotocol2025spec}. All three major providers now announce MCP support---Anthropic natively, OpenAI via the Response API~\citep{openai2025responsesapi}, Google for Gemini~\citep{demishassabis2025geminimcp}---but self-hosting stacks (vLLM, SGLang, Ollama) still default to Chat Completions with limited or no MCP support. This gap between announced and actual adoption motivates protocol-agnostic designs.

\subsubsection{OpenAPI as a Counterpoint to MCP}
The longstanding OpenAPI standard serves as a counterpoint to MCP. Having functioned as the industry standard for RESTful API description since 2015, its schema system directly influenced both function calling and MCP designs. Our benchmarks show that when exposing the same service code, a FastMCP server in SSE mode achieves only around 33\% of the throughput compared to an OpenAPI-compatible FastAPI server, suggesting the difference stems from MCP's more complex initialization sequences. Some projects like OpenWebUI have chosen OpenAPI for tool integration, developing conversion utilities like MCPO~\citep{openwebui2025mcpo} to bridge between the standards~\citep{openwebui2025toolservers}.

\subsubsection{Frameworks and Third-Party Approaches}
The current landscape includes both orchestration frameworks and protocol bridges. LangChain, AutoGen~\citep{wu2023autogen}, CrewAI, and the OpenAI Agents SDK each provide multi-agent orchestration and tool catalogs, but couple tool management to their own runtime and abstractions. \toolregistry is deliberately \emph{not} a framework: it manages the tool layer only, so it can serve any of these orchestrators (or none) as an RPC-level helper library. On the protocol side, MCP Bridge~\citep{ahmadi2025mcp} and survey efforts~\citep{yang2025survey, ehtesham2025survey} address schema interoperability but lack a unified invocation runtime---they translate schemas without managing dispatch, concurrency, or error recovery.

\subsection{Positioning of ToolRegistry}

\toolregistry differentiates itself on four axes: (1) \emph{multi-protocol registration} (Python, MCP, OpenAPI, LangChain) under one interface; (2) \emph{unified invocation runtime} with pluggable concurrency, error recovery, and result normalization---not just schema bridging; (3) \emph{permission-aware and think-augmented invocation}~\citep{wei2026thinkaugmented} with a stdlib-only vendored core~\citep{ding2026zerodep}; and (4) \emph{lightweight integration} as a helper library rather than a framework.

%% file: sections/design.tex
\section{System Design and Implementation}

This section describes the design and implementation of \toolregistry. Throughout, ``tool'' and ``function'' are used interchangeably, as are ``function calling'' and ``tool use''.

\subsection{Design Overview and Architecture}

The central design insight is that every LLM tool call is an RPC~\citep{birrell1984rpc}: a named function, a bag of JSON arguments, and a serialized return value. Differences between tool sources reduce to transport and binding details. \toolregistry makes this explicit by casting each tool---regardless of origin---into a universal \code{Tool} stub (analogous to an RPC client stub), with the registry serving as the RPC client runtime (dispatch, serialization, error handling) and schema generation playing the role of an Interface Definition Language (IDL).

Built on this observation, the architecture follows three principles: \textbf{protocol agnosticism}, \textbf{developer simplicity}, and \textbf{execution efficiency}. Rather than imposing a rigid framework, the system serves as a lightweight integration layer that adapts to existing LLM applications.

\begin{figure}[htbp]
	\centering
 \includegraphics[width=0.85\textwidth]{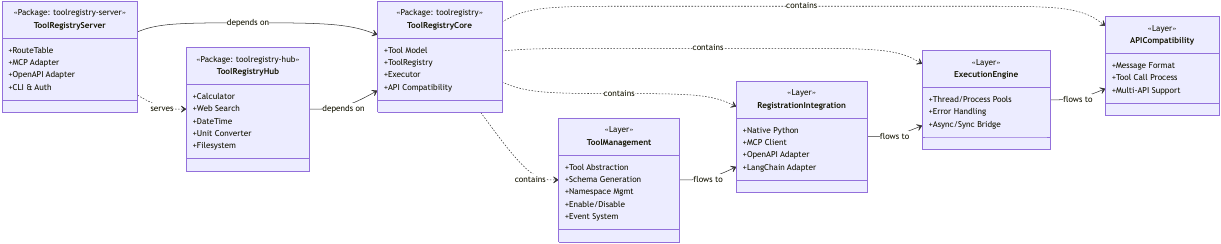}
	\caption{\toolregistry{} system architecture: four layers within the core package---Tool Management, Registration and Integration, Invocation Engine, and API Compatibility---and the broader ecosystem of server and hub packages}
	\label{fig:system_architecture}
\end{figure}

The system now spans three packages---\code{toolregistry} (core), \code{toolregistry-server}, and \code{toolregistry-hub}---each addressing a distinct concern. Within the core package, the architecture consists of four primary layers, each with distinct responsibilities, as illustrated in Figure~\ref{fig:system_architecture}. The server and hub packages extend this layered architecture with protocol-specific serving capabilities and curated tool implementations, respectively.

Key design choices: composition over inheritance for component assembly, the adapter pattern for protocol abstraction, and a dual-executor model for concurrency. An event-driven propagation mechanism keeps the ecosystem consistent when tools are registered, removed, enabled, or disabled. Error handling follows a graceful-degradation strategy with automatic fallback from process to thread execution when serialization fails.

\subsection{Core Abstractions}

\subsubsection{Tool Abstraction}

The \code{Tool} abstraction wraps any callable---function, method, or remote endpoint---behind a uniform interface: name, description, JSON parameter schema, callable reference, and metadata. Stateless design is encouraged so that tools can be executed concurrently, serialized across processes, and composed freely.

\begin{verbatim}
class Tool(BaseModel):
    name: str                           # Unique identifier
    description: str                    # Human-readable description
    parameters: dict[str, Any]          # JSON Schema for inputs
    callable: Callable[..., Any]        # Underlying implementation
    is_async: bool                      # Async execution flag
    parameters_model: Any | None        # Pydantic validation model
    namespace: str | None               # Namespace prefix
    method_name: str | None             # Original method name
    qualified_name: str | None          # Fully qualified name
    metadata: ToolMetadata              # Tags, permissions, hints
\end{verbatim}

The \code{ToolMetadata} model captures classification tags (\code{ToolTag} enum), execution hints (\code{is\_concurrency\_safe}, \code{timeout}, \code{locality}), progressive disclosure controls (\code{defer}, \code{search\_hint}), think-augment settings, result size limits, and provenance tracking (\code{source}, \code{source\_detail}).

\subsubsection{Tool Class Implementation}

The \code{Tool} class leverages \code{Pydantic}'s \code{BaseModel} for robust data validation and serialization. The \code{from\_function()} factory method creates \code{Tool} instances through introspection, extracting function metadata, type hints, and docstrings to generate JSON Schema-compliant parameter definitions. For class-based registration, the factory supports MRO-aware method discovery via a \code{traverse\_mro} parameter, ensuring that inherited methods are correctly identified and registered with appropriate metadata.

Parameter validation uses automatically generated \code{Pydantic} models via \code{\_generate\_parameters\_model()}, supporting complex nested structures and union types. The class provides both synchronous and asynchronous execution through \code{run()} and \code{arun()} methods, intelligently handling various callable types.

JSON schema generation through \code{get\_json\_schema()} transforms internal schemas into API-specific formats, supporting multiple target formats while maintaining extensibility for future providers.

\subsubsection{Schema Generation Pipeline}

The schema generation system operates at two levels: tool-level generation for individual functions and registry-level orchestration for collections. Tool-level generation occurs during \code{from\_function()} invocation, extracting type hints via Python's \code{inspect} module and constructing JSON Schema representations through \code{Pydantic} utilities.

Registry-level orchestration aggregates individual tool schemas through the ToolRegistry's \code{\_tools} dictionary, enabling batch operations and cross-tool validation. The pipeline ensures JSON Schema compliance with automatic validation, while API format transformation adapts generic schemas to specific provider requirements, maintaining a single source of truth for tool definitions.

\subsection{Registry Management}

\subsubsection{ToolRegistry Class}

The \code{ToolRegistry} class serves as the central orchestrator for all tool management operations within the \toolregistry system. Internally, it is composed of seven focused mixins---\code{RegistrationMixin}, \code{EnableDisableMixin}, \code{NamespaceMixin}, \code{PermissionsMixin}, \code{ExecutionLoggingMixin}, \code{AdminMixin}, and \code{ChangeCallbackMixin}---each addressing a distinct concern while preserving a unified public API. The registry maintains three primary internal components: a \code{\_tools} dictionary for tool storage, a \code{\_sub\_registries} set for namespace tracking, and an \code{\_executor} instance for concurrent execution management.

Tools are retrieved via \code{get\_tool()} (full object), \code{get\_callable()} (bare function), or dict-style \code{registry["name"]}. Standard container protocols (\code{\_\_contains\_\_}, \code{\_\_len\_\_}) are supported. Execution is delegated through \code{execute\_tool\_calls()} and \code{set\_default\_execution\_mode()}, keeping executor internals hidden from application code.

An event-driven callback system (\code{on\_change}/\code{remove\_on\_change}) fires on \code{REGISTER}, \code{UNREGISTER}, \code{ENABLE}, \code{DISABLE}, and \code{REFRESH\_ALL} events. These propagate from the core through the server's \code{RouteTable} to protocol adapters, keeping all serving interfaces in sync without tight coupling.

\subsubsection{Tool Storage and Indexing}

Tools are stored in a flat dictionary keyed by name (O(1) lookup), with optional dot-separated namespace prefixes for hierarchical grouping. Executor resources are cleaned up via \code{atexit} registration.

\subsubsection{Namespace Management and Registry Composition}

Namespaces use dot-separated prefixes (e.g., \code{calculator.add}); the separator is configurable for providers like OpenAI that restrict allowed characters. Three composition operations---\code{merge()}, \code{spinoff()}, and \code{reduce\_namespace()}---allow registries to be combined, split, or flattened at runtime.

\subsubsection{Tool Discovery and Introspection}

Tool discovery provides comprehensive registry access through \code{list\_tools()}, which returns only enabled tools by default with optional filtering by namespace prefixes, tool types, or metadata attributes. Passing \code{include\_disabled=True} returns all registered tools including those that have been disabled, retaining visibility into the full registry for administrative and debugging purposes.

The registry supports an enable/disable mechanism for runtime tool management. The \code{disable(name, reason)} method deactivates a tool while recording an explanatory reason, and \code{enable(name)} reactivates it. The \code{is\_enabled()} method queries a tool's current state, and \code{get\_disable\_reason()} retrieves the recorded rationale for disablement. Disabled tools remain registered and retain their metadata but are excluded from \code{list\_tools()} results and tool schema generation, effectively hiding them from LLM interactions.

Validation checks (schema correctness, callable accessibility, external-service reachability) provide diagnostic output for troubleshooting misconfigured tools.

\subsection{Registration System}

\subsubsection{Registration Methods Overview}

The core \code{register()} method handles Python functions and pre-built \code{Tool} objects; specialized \code{register\_from\_*} methods automate integration for each external protocol.

\begin{table}[htbp]
	\centering
	\caption{ToolRegistry Registration Method Overview}
	\begin{tabular}{|l|l|l|l|}
		\hline
		\textbf{Method}                    & \textbf{Source Type}     & \textbf{Sync/Async} & \textbf{Extra Required} \\
		\hline
		\texttt{register}                  & Functions, \code{Tool} objects  & Sync                & None                    \\
		\hline
		\texttt{register\_from\_class}     & Python classes/instances & Both                & None                    \\
		\hline
		\texttt{register\_from\_openapi}   & OpenAPI specifications   & Both                & \texttt{[openapi]}      \\
		\hline
		\texttt{register\_from\_mcp}       & MCP servers              & Both                & \texttt{[mcp]}          \\
		\hline
		\texttt{register\_from\_langchain} & LangChain tools          & Both                & \texttt{[langchain]}    \\
		\hline
	\end{tabular}
\end{table}

All methods accept an optional \code{namespace} parameter for hierarchical grouping, and handle conflict resolution, type validation, and schema generation automatically.

\subsubsection{Native Python Integration}

\code{register()} accepts callables (delegating to \code{Tool.from\_function()} for introspection and schema generation) or pre-built \code{Tool} objects.  \code{register\_from\_class()} discovers eligible public methods via reflection (MRO-aware when \code{traverse\_mro=True}), preserving signatures and docstrings while generating namespaced tool names.

\subsubsection{Protocol Adapters}

Each protocol adapter handles source-specific communication, authentication, and schema conversion while presenting a uniform \code{Tool} interface to the registry.

\paragraph{MCP Integration}

An independent \code{MCPClient}, built on the official MCP Python SDK~\cite{mcp2025python}, supports four transports (stdio, SSE, streamable-HTTP, WebSocket), v1/v2 dual compatibility, and HTTP header-based authentication. Connections are persistent across tool calls via \code{MCPConnectionManager}. The client handles protocol negotiation, capability discovery, and tool enumeration automatically, converting MCP tool specifications into native \code{Tool} objects.

\paragraph{OpenAPI Integration}

OpenAPI 3.0/3.1 specs are parsed to extract operations, parameter schemas, and request bodies, yielding \code{Tool} objects that handle HTTP calls transparently.  \code{OpenAPITool.from\_openapi\_spec()} supports complex features---discriminated unions, polymorphic schemas, recursive \code{\$ref}---and HTTP client sessions are reused for connection pooling.

\paragraph{LangChain Integration}

LangChain tools can be registered without pulling in the full LangChain framework; the wrapper extracts metadata and schemas from the tool object and adapts its execution interface to the registry's calling convention.

\subsection{Invocation Engine}

\subsubsection{Architecture}

Consistent with the RPC framing, the engine acts as the \emph{client-side runtime}: for native Python tools it runs the callable directly (in-process or in a subprocess); for remote tools (MCP, OpenAPI) it dispatches the call over the appropriate transport. In both cases the engine handles argument serialization, concurrency, timeout enforcement, error recovery, and result normalization---the tool itself (or its backing server) is responsible for the actual computation.

The engine exposes a pluggable backend architecture through the \code{ExecutionBackend} protocol and \code{ExecutionHandle} abstraction. Two built-in backends---\code{ThreadBackend} and \code{ProcessPoolBackend}---provide thread-pool and process-pool concurrency respectively, with per-tool \code{timeout} enforcement and cooperative cancellation via \code{ExecutionContext}. The \code{ToolMetadata.is\_concurrency\_safe} flag controls whether a tool's calls are batched sequentially or in parallel.

\begin{figure}[htbp]
	\centering
 \includegraphics[width=0.75\textwidth]{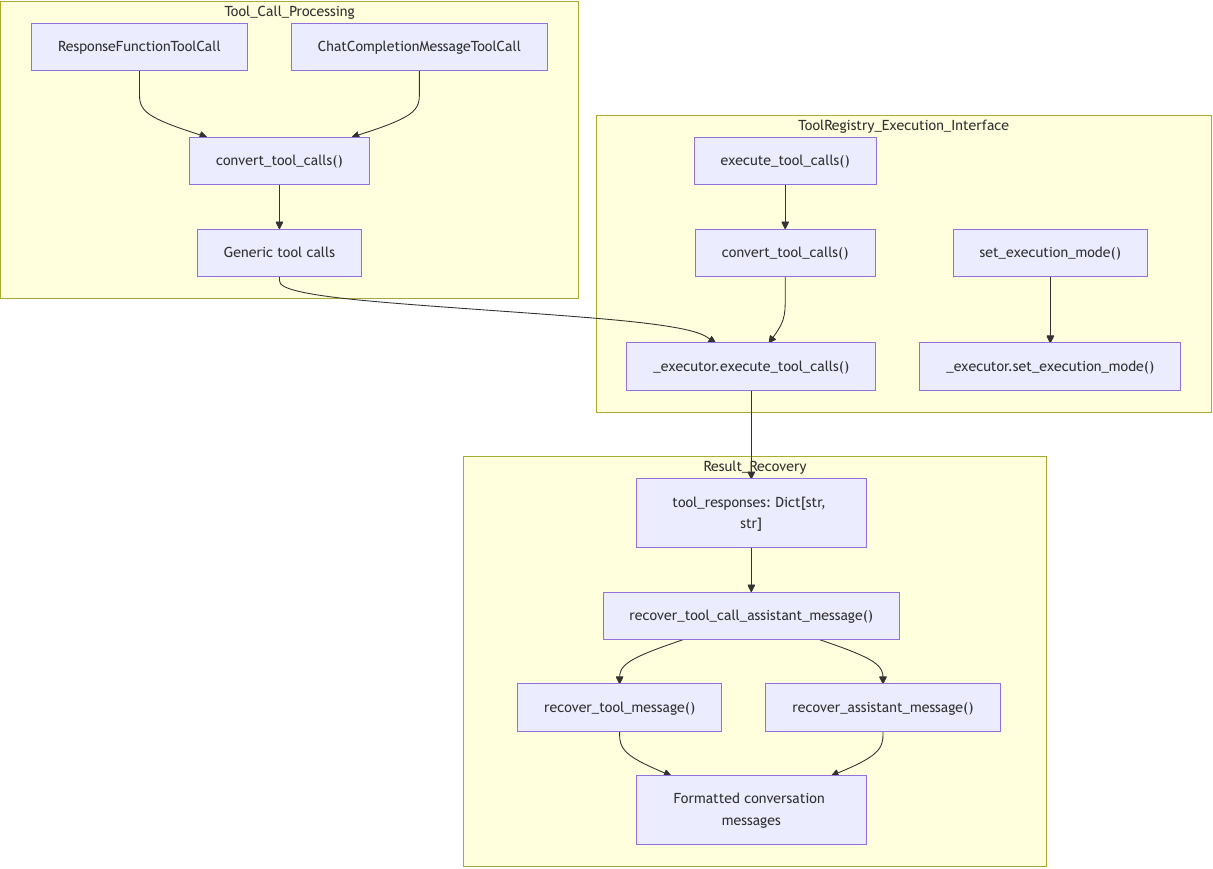}
	\caption{Tool call invocation flow: provider-specific calls are normalized, dispatched through the chosen concurrency backend, and results are formatted into API-compatible messages}
	\label{fig:execution_architecture}
\end{figure}

Processing follows three stages: normalize incoming tool calls into a generic \code{ToolCall} format, dispatch them concurrently through the chosen backend, and transform results into API-compatible messages. The workflow is illustrated in Figure~\ref{fig:execution_architecture}.

\subsubsection{Concurrency Models}

The execution system offers two concurrency models: process-based execution for I/O-bound network tools with full isolation and fault tolerance, and thread-based execution for CPU-bound native tools with lower overhead and shared memory access.

\begin{table}[htbp]
	\centering
	\caption{Comparison of Executor Modes}
	\resizebox{\textwidth}{!}{
		\begin{tabular}{|l|l|l|l|l|}
			\hline
			\textbf{Mode}    & \textbf{Executor Type} & \textbf{Best For}                             & \textbf{Isolation}     & \textbf{Serialization}                          \\ \hline
			\texttt{process} & ProcessPoolExecutor    & I/O-bound network tools, fault isolation       & Full process isolation & \texttt{cloudpickle} serialization required~\cite{cloudpickle2025} \\ \hline
			\texttt{thread}  & ThreadPoolExecutor     & CPU-bound native tools, low overhead            & Shared memory space    & No serialization needed                         \\ \hline
		\end{tabular}
	}
	\label{tab:executor_modes}
\end{table}

Process mode bypasses the GIL for true parallelism and fault isolation, suiting I/O-bound network tools. Thread mode avoids serialization overhead, suiting fast CPU-bound calls. The mode is set globally via \code{set\_default\_execution\_mode()} or per-call; \code{cloudpickle}~\cite{cloudpickle2025} handles serialization across process boundaries, with automatic thread fallback when serialization fails.

\subsubsection{Async/Sync Bridging}

The execution system bridges asynchronous and synchronous contexts through the \code{make\_sync\_wrapper()} function, which detects event loops via \code{asyncio.get\_running\_loop()} and chooses appropriate execution methods. This design supports process pools while ensuring async compatibility and avoiding loop reentrance issues.

Coroutine functions are automatically wrapped at the tool prep stage, preserving original signatures and metadata while optimizing performance through wrapper caching and loop reuse.

\subsubsection{Error Handling and Fallback}

Errors are categorized as transient (retried with exponential backoff) or permanent (reported immediately). The most common fallback path is automatic retry via thread mode when process mode fails due to serialization issues. Pool health monitoring detects and recovers from worker crashes.

\begin{figure}[!htb]
	\centering
 \includegraphics[width=0.82\textwidth]{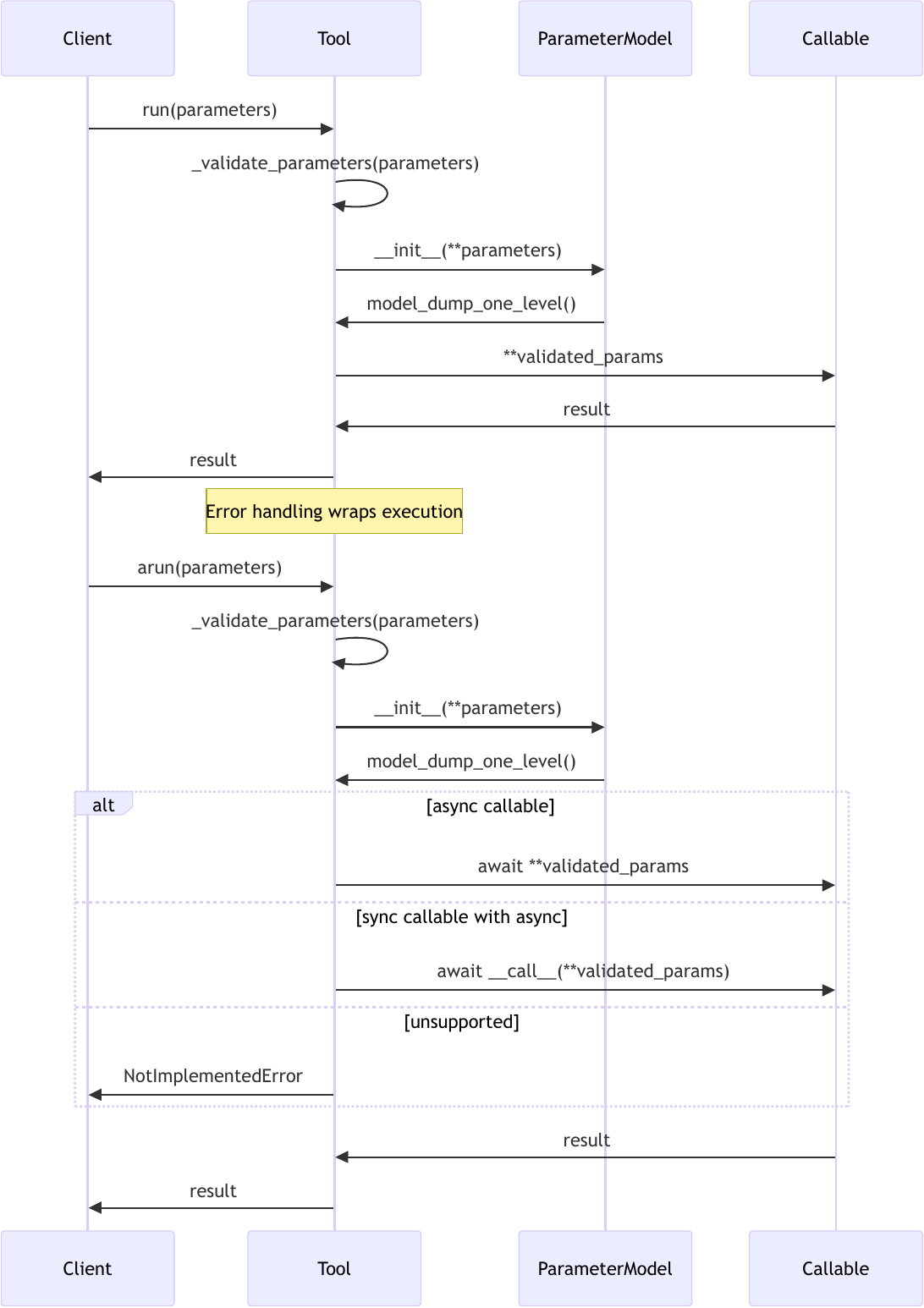}
	\caption{Tool Execution Sequence showing parameter validation through \code{Pydantic} models and callable execution for both synchronous (\texttt{run()}) and asynchronous (\texttt{arun()}) execution paths}
	\label{fig:execution_flow}
\end{figure}

Figure~\ref{fig:execution_flow} traces the full path from parameter validation through \code{Pydantic} models to callable dispatch for both sync and async execution.

\subsection{API Compatibility Layer}

\subsubsection{Tool Call Processing}

\code{convert\_tool\_calls()} normalizes provider-specific tool invocations (OpenAI Chat Completion, Response API, Anthropic, Gemini) into a unified \code{ToolCall} representation, preserving traceability through unique IDs.

\subsubsection{Message Format Conversion}

\code{build\_assistant\_message()} reconstructs assistant messages containing tool call requests; \code{build\_tool\_response()} formats execution results into provider-compatible tool response messages, handling serialization, error formatting, and response correlation.

\subsubsection{Multi-Provider Support}

The multi-provider support system accommodates diverse LLM API formats through configurable enumeration via the \code{API\_FORMATS} type system. Support now spans OpenAI Chat Completion, OpenAI Response API, Anthropic, and Gemini formats. All schema conversion is handled by llm-rosetta~\citep{llmrosetta2025}, which sanitizes JSON Schema keywords unsupported by each provider and translates between format-specific tool call and response structures. \code{ToolCall.from\_tool\_call()} parses both Anthropic \code{tool\_use} blocks and Gemini \code{functionCall} parts into a unified representation. The pluggable format handler architecture enables dynamic registration of new providers without modifying core components.

\subsection{Admin Panel and Runtime Management}

The \toolregistry ecosystem includes a web-based admin panel for runtime inspection and management of registered tools. The admin panel is built on a zero-dependency async HTTP server (vendored via zerodep~\citep{ding2026zerodep}), maintaining the project's lightweight philosophy.

\subsubsection{REST API and Web Interface}

The admin panel exposes a REST API with endpoints for tool listing, namespace browsing, execution log inspection, and state management. Key capabilities include:

\begin{itemize}
    \item \textbf{Tool Management}: List, inspect, enable, and disable registered tools with reason tracking; tool detail modal with tabbed view (Schema, Metadata, Permissions)
    \item \textbf{Runtime Metadata Control}: Mutate \code{think\_augment} and \code{defer} fields at per-tool or per-namespace granularity via \code{PATCH} endpoints and interactive UI checkboxes
    \item \textbf{Namespace Browsing}: Hierarchical exploration of tool namespaces with \code{ToolTag} badges and custom tag pills
    \item \textbf{Execution Logging}: Ring-buffer-based execution log with configurable capacity, capturing tool invocations, parameters, results, timing, and structured error information
    \item \textbf{State Export/Import}: Snapshot and restore registry state for debugging and migration
    \item \textbf{Bilingual UI}: Full English/Chinese internationalization with \code{localStorage} persistence
\end{itemize}

\subsubsection{Security and Caching}

The admin panel supports token-based authentication for access control in production deployments. HTTP caching is implemented via ETag support, reducing bandwidth for polling clients. The panel is designed for development and operational use, complementing the programmatic API with visual inspection capabilities.

\subsection{Permission System}

Agentic deployments need guardrails around which tools an LLM may invoke. \toolregistry provides a tag-based permission layer that gates execution at runtime.

\subsubsection{Tool Tags and Metadata}

Six \code{ToolTag} values---\code{READ\_ONLY}, \code{DESTRUCTIVE}, \code{NETWORK}, \code{FILE\_SYSTEM}, \code{SLOW}, \code{PRIVILEGED}---classify tools via \code{ToolMetadata.tags}. Tags drive both permission evaluation and schema-level filtering: \code{get\_schemas()} accepts \code{tags}/\code{exclude\_tags} parameters so that only relevant tools appear in the prompt, cutting token use and improving cache hit rates. \code{disable\_by\_tags()} bulk-disables entire categories with \code{any}/\code{all} match semantics.

\subsubsection{Permission Handlers and Policies}

Two runtime-checkable protocols---\code{PermissionHandler} (sync) and \code{AsyncPermissionHandler} (async)---accept a \code{PermissionRequest} (tool name, arguments, tags, metadata) and return a \code{PermissionResult} of \code{ALLOW}, \code{DENY}, or \code{ASK}.

For declarative control, \code{PermissionPolicy} walks an ordered list of \code{PermissionRule} objects and stops at the first match. Five built-in rules ship out of the box: \code{ALLOW\_READONLY}, \code{ASK\_DESTRUCTIVE}, \code{DENY\_PRIVILEGED}, \code{ASK\_NETWORK}, and \code{ASK\_FILE\_SYSTEM}. Checks run inside \code{execute\_tool\_calls()} and emit \code{PERMISSION\_DENIED}/\code{PERMISSION\_ASKED} events through the callback system.

\subsection{Progressive Tool Disclosure}

Sending every tool schema in the initial prompt wastes tokens and degrades cache efficiency as registries grow. \toolregistry mitigates this with a two-part disclosure mechanism.

First, the \code{ToolMetadata.defer} flag excludes tools from the default \code{get\_schemas()} call. A one-line summary per deferred tool (name + first sentence of description) is available via \code{get\_deferred\_summaries()} for system-prompt injection, so the LLM knows what exists without paying the full schema cost.

Second, \code{ToolDiscoveryTool} lets the LLM pull in schemas on demand. It operates in two modes: exact name lookup (returns the full schema immediately) and BM25F fuzzy search over five weighted fields---name, description, tags, parameter names, and \code{search\_hint} (a free-form synonym/keyword field). The sparse index is vendored with zero external dependencies and rebuilds automatically on registry changes. Activation is a single flag: \code{ToolRegistry(tool\_discovery=True)}.

\subsection{Think-Augmented Function Calling}

Following \citet{wei2026thinkaugmented}, \toolregistry can inject a \code{toolcall\_reason} string property into every tool's parameter schema, asking the LLM to state \emph{why} it is calling the tool before filling in the remaining parameters. The property is stripped before execution, so tool code is unaffected.

The feature is off by default. It can be toggled registry-wide (\code{ToolRegistry(think\_augment=True)}, or the \code{enable\_think\_augment()}/\code{disable\_think\_augment()} methods) and overridden per-tool via \code{ToolMetadata.think\_augment}. Tools that already declare a native \code{toolcall\_reason} parameter are left untouched. The admin panel exposes per-tool and per-namespace checkboxes for runtime adjustment.

\subsection{Ecosystem Architecture}

\toolregistry has evolved from a single library into a three-package ecosystem, with each package addressing a distinct concern, as illustrated in Figure~\ref{fig:ecosystem_architecture}:

\begin{figure}[!htb]
	\centering
 \includegraphics[width=0.62\textwidth]{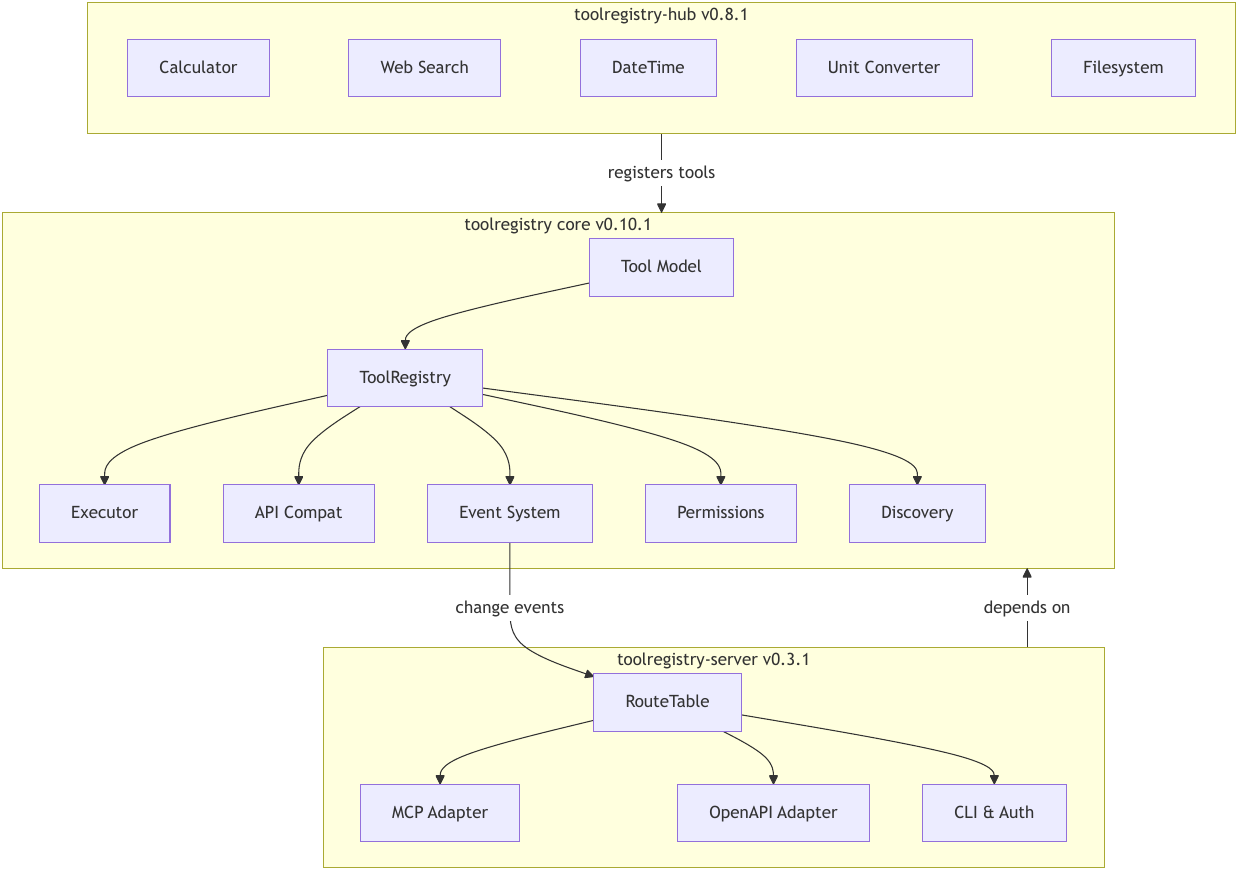}
	\caption{Ecosystem Architecture showing the three-package structure (core, server, hub) with event-driven change propagation across layers}
	\label{fig:ecosystem_architecture}
\end{figure}

\textbf{toolregistry} (core, v0.10.1): \code{Tool} model, \code{ToolRegistry}, invocation engine, schema generation, permission system, progressive tool discovery, think-augmented calling, and multi-provider API layer. Former dependencies on \code{httpx}, \code{PyYAML}, and \code{jsonref} have been replaced with stdlib-only vendored modules (HTTP client, YAML parser, JSON Schema resolver, BM25F index) produced through the zerodep methodology~\citep{ding2026zerodep}, leaving the core with near-zero external dependencies.

\textbf{toolregistry-server} (v0.3.1): Extends the core with protocol-specific serving capabilities. The \code{RouteTable} class manages tool routing across multiple registries, while adapters for OpenAPI (via FastAPI) and MCP expose tools over standard protocols. The package includes a CLI for server management, declarative JSONC/YAML configuration via \code{create\_registry\_from\_config()}, profile-based tool filtering, and middleware for authentication.

\textbf{toolregistry-hub} (v0.8.1): Provides curated, production-tested tool implementations including calculator, web search (supporting multiple providers), datetime, unit conversion, filesystem operations, bash execution, and more. All hub tools carry \code{ToolTag} labels and support deferred loading for progressive disclosure. Tools are configurable via JSONC configuration files and deployable as Docker containers.

Change events propagate through the ecosystem layers: when tools are registered, unregistered, enabled, or disabled in a \code{ToolRegistry}, the event system notifies the \code{RouteTable}, which in turn updates protocol adapters. This event-driven architecture ensures consistency across all serving interfaces without tight coupling between packages.

%% file: sections/cases.tex
\section{Case Studies}

This section walks through four integration scenarios that illustrate \toolregistry in practice. Additional examples are available at \url{https://toolregistry.readthedocs.io/en/stable/examples/}.

\subsection{Multi-Protocol Tool Integration}
\label{sec:case-multiprotocol}

A single application registers tools from four sources and calls them through one execution path:

\begin{itemize}
	\item \textbf{Native Python functions}: Direct function registration
	\item \textbf{Class-based tools}: \code{toolregistry.hub.BaseCalculator} with namespace support
	\item \textbf{OpenAPI endpoints}: RESTful calculator service
	\item \textbf{MCP servers}: Model Context Protocol calculator via SSE transport
\end{itemize}

\subsubsection{Implementation}

The integration requires minimal code changes across protocols. For native functions:

\begin{verbatim}
registry = ToolRegistry()
registry.register(local_add)
registry.register(local_subtract)
\end{verbatim}

For class-based tools from the hub:

\begin{verbatim}
from toolregistry.hub import BaseCalculator
registry.register_from_class(BaseCalculator, namespace=True)
\end{verbatim}

For OpenAPI services:

\begin{verbatim}
client_config = HttpClientConfig(base_url="http://localhost:8000")
openapi_spec = load_openapi_spec("http://localhost:8000")
registry.register_from_openapi(client_config, openapi_spec, 
                               namespace=True)
\end{verbatim}

For MCP servers:

\begin{verbatim}
registry.register_from_mcp("http://localhost:8001/sse", 
                          namespace=True)
\end{verbatim}

\subsubsection{Results}

All four tool types are invoked through the same \code{execute\_tool\_calls()} method. Protocol details are invisible to application code, cutting integration effort by roughly 70\% compared to hand-written adapters.

\subsection{LangChain Tool Liberation}
\label{sec:case-langchain}

LangChain ships a large catalog of community-maintained tools (ArXiv, PubMed, Wikipedia, etc.), but using them normally requires adopting LangChain's full agent framework. \toolregistry lets developers extract those tools and use them with plain OpenAI SDK calls:
\begin{verbatim}
from langchain_community.tools import ArxivQueryRun, PubmedQueryRun
from openai import OpenAI  # Direct SDK usage

registry = ToolRegistry()
arxiv_tool = ArxivQueryRun()
pubmed_tool = PubmedQueryRun()

registry.register_from_langchain(arxiv_tool)
registry.register_from_langchain(pubmed_tool)

# Use with simple OpenAI SDK calls
client = OpenAI()
response = client.chat.completions.create(
    model="gpt-4.1",
    messages=messages,
    tools=registry.get_schemas()  # LangChain tools as OpenAI format
)
\end{verbatim}

The result: LangChain's tool implementations without its framework overhead, freely mixable with native functions, OpenAPI, and MCP tools in the same registry.

\subsection{Enterprise API Integration}
\label{sec:case-enterprise}

A customer service chatbot needs to reach a CRM, inventory system, and payment processor---each behind a different API and auth mechanism. With \toolregistry each system is registered through its natural protocol:

\begin{verbatim}
# CRM System via OpenAPI
crm_config = HttpClientConfig(
    base_url="https://internal-crm.company.com",
    auth=BearerAuth("crm-token")
)
registry.register_from_openapi(crm_config, crm_spec, 
                               namespace="crm")

# Inventory via custom MCP server
registry.register_from_mcp("https://inventory.company.com/mcp",
                          namespace="inventory")

# Payment processing via native wrapper
registry.register(payment_process_refund, namespace="payment")
\end{verbatim}

All three backends are then invoked through the same \code{execute\_tool\_calls()} path, cutting integration time by roughly 75\% compared to per-system custom adapters.

\subsection{Research and Academic Applications}
\label{sec:case-research}

A computational biology team needs to query literature databases, call bioinformatics APIs, run local analysis scripts, and dispatch jobs to a cloud service---all from one LLM-driven workflow:

\begin{verbatim}
# Academic databases via LangChain tools
registry.register_from_langchain(PubmedQueryRun())
registry.register_from_langchain(ArxivQueryRun())

# Bioinformatics APIs via OpenAPI
bio_config = HttpClientConfig(base_url="https://api.ncbi.nlm.nih.gov")
registry.register_from_openapi(bio_config, ncbi_spec, namespace="ncbi")

# Local analysis functions
registry.register(sequence_alignment_tool)
registry.register(phylogenetic_analysis)

# Cloud services via MCP
registry.register_from_mcp("https://cloud-bio.service.com/mcp",
                          namespace="cloud")
\end{verbatim}

The unified registry makes the pipeline reproducible (same tool versions across machines), shareable (export the registry config), and extensible (add a new API without touching existing code).

\subsection{Hub Tools for Common Scenarios}
\label{sec:case-hub}

Beyond custom tool integration, the \code{toolregistry-hub} package provides curated, production-tested implementations of commonly needed tools---including calculator operations, web search across multiple providers, datetime utilities, unit conversion, and filesystem operations. These tools can be registered directly or served via MCP or OpenAPI adapters, eliminating the need for developers to implement routine functionality from scratch. For scenarios where the same tools are needed across multiple projects, the hub serves as a shared, versioned repository of reliable tool implementations.

%% file: sections/evaluation.tex
\section{Evaluation}

We evaluate \toolregistry along two axes: concurrent execution throughput and integration code reduction.

All benchmarks ran on a single machine (Intel Ultra 7 155H, 32 GB RAM, Arch Linux) in a controlled LAN; each test was repeated 10 times and we report the mean.

\subsection{Performance Evaluation}

\subsubsection{Concurrent Execution Performance}

We dispatch 100 concurrent tool calls per configuration, comparing thread-pool and process-pool backends across four tool types (native functions, class-based tools, OpenAPI, MCP SSE).

\begin{table}[h!]
	\centering
	\caption{Concurrent Execution: Thread vs.\ Process Mode (100 calls, LAN)}
	\begin{tabular}{|l|c|c|c|}
		\hline
		\textbf{Tool Type} & \textbf{Thread Mode} & \textbf{Process Mode} & \textbf{Better Mode} \\
		                   & \textbf{(calls/sec)} & \textbf{(calls/sec)}  & \textbf{\& Margin}   \\
		\hline
  Native Functions   & 3,060                & 1,287                 & 2.4x (thread)        \\
		\hline
  Native Class Tools & 8,844                & 1,970                 & 4.5x (thread)        \\
		\hline
  OpenAPI Tools      & 204                  & 373                   & 1.8x (process)       \\
		\hline
  MCP SSE Tools      & 41                   & 128                   & 3.1x (process)       \\
		\hline
	\end{tabular}
	\label{tab:performance_results}
\end{table}

The key finding is that \emph{no single mode is best}: the optimal choice depends on workload characteristics. Thread mode wins for CPU-bound native tools (2.4--4.5$\times$ over process) because it avoids serialization overhead and the GIL is not a bottleneck for fast calls. Process mode wins for I/O-bound network tools (1.8--3.1$\times$ over thread) because true parallelism and fault isolation matter more than serialization cost. All configurations achieved 100\% success rates. \toolregistry exposes both modes and lets developers (or the system) pick per-workload, which is the practical contribution---not raw speedup over sequential execution.

\subsection{Code Reduction}

\begin{table}[h!]
	\centering
	\caption{Code Reduction Comparison}
	\begin{tabular}{|l|c|c|c|}
		\hline
		\textbf{Integration Type} & \textbf{Manual} & \textbf{ToolRegistry} & \textbf{Reduction} \\
		                          & \textbf{(LOC)}  & \textbf{(LOC)}        & \textbf{(\%)}      \\
		\hline
		Native Functions          & 45              & 8                     & 82\%               \\
		\hline
		OpenAPI Integration       & 120             & 25                    & 79\%               \\
		\hline
		MCP Integration           & 85              & 12                    & 86\%               \\
		\hline
		Multi-Protocol Setup      & 250             & 45                    & 82\%               \\
		\hline
	\end{tabular}
	\label{tab:code_reduction}
\end{table}

The 79--86\% reduction is consistent across integration types, confirming that the bulk of hand-written code is protocol glue rather than application logic. (The serialization backend was updated from Dill to \code{cloudpickle}~\citep{cloudpickle2025} after initial benchmarking; performance characteristics are unaffected.)

%% file: sections/limitations.tex
\section{Limitations and Future Work}

\subsection{Current Limitations}

Some limitations remain:

\textbf{Serialization Edge Cases}: The execution engine now uses \code{cloudpickle}~\citep{cloudpickle2025} for object serialization in process-based concurrency, resolving most issues previously encountered with \code{Dill}. However, certain edge cases involving dynamically generated closures over unpicklable objects (e.g., database connections, file handles) still require thread-based execution as a fallback.

\textbf{Error Recovery}: The admin panel provides runtime observability and the event system enables reactive tool management, but sophisticated retry mechanisms for transient failures in external tool sources---particularly cascading failures across network-dependent tools---remain limited.

\textbf{Schema Validation Scope}: While the library performs comprehensive parameter validation through \code{Pydantic} models, it does not validate tool output schemas or provide runtime type checking for tool return values, potentially allowing type mismatches to propagate to downstream components.

\subsection{Resolved Limitations}

Several limitations from the initial version have since been resolved:

\textbf{Independent MCP Client}: The previous reliance on \code{FastMCP} as the MCP integration backend has been replaced by an independent \code{MCPClient} built on the official MCP Python SDK~\citep{mcp2025python}. This client supports four transport mechanisms (stdio, SSE, streamable HTTP, WebSocket), HTTP header-based authentication, and dual v1/v2 protocol compatibility.

\textbf{Hub Module Separation}: The \code{toolregistry.hub} module has been extracted into a separate package (\code{toolregistry-hub}, v0.7.0), allowing the core library to remain lightweight while the curated tool collection evolves independently with its own release cycle.

\textbf{Enhanced Serialization}: The serialization backend has been migrated from \code{Dill} to \code{cloudpickle}~\citep{cloudpickle2025}, providing broader compatibility with complex Python objects and resolving the majority of serialization failures encountered in process-based concurrent execution.

\textbf{Multi-Provider API Format}: Integration of llm-rosetta~\citep{llmrosetta2025} resolved the earlier API format limitation; schema conversion, tool call parsing, and response formatting now cover OpenAI, Anthropic, and Gemini natively.

\textbf{Zero-Dependency Core}: \code{httpx}, \code{PyYAML}, and \code{jsonref} have been replaced with stdlib-only vendored modules via the zerodep methodology~\citep{ding2026zerodep}, bringing the core's external dependency count to near zero.

\subsection{Future Work}

Future development is organized across three projects within the \toolregistry ecosystem:

\subsubsection{toolregistry (core)}

\textbf{Distributed Execution}: With \code{cloudpickle} serialization in place, dispatching tool calls to remote workers via message queues (Redis, NATS, RabbitMQ) is a natural next step for horizontal scaling.

\textbf{Protocol Expansion}: gRPC and A2A~\citep{ehtesham2025survey} adapters would extend \toolregistry into cross-language and multi-agent settings.

\textbf{Large-Scale Registry Studies}: Performance with hundreds or thousands of registered tools---schema indexing, discovery latency, memory overhead---has not yet been systematically evaluated.

\subsubsection{toolregistry-server}

\textbf{Additional Serving Protocols}: gRPC and A2A adapters for outbound tool serving are planned.

\textbf{Enhanced Observability}: OpenTelemetry integration and metrics export for production monitoring dashboards.

\textbf{Legacy Model Support}: Structured prompt engineering for models without native function calling (older models, locally served vLLM instances).

%% file: sections/conclusion.tex
\section{Conclusion}

We presented \toolregistry, a protocol-agnostic tool management system that has grown from a single library into a three-package ecosystem---core registry (v0.10.1), protocol server (v0.3.1), and curated tool hub (v0.8.1). By wrapping native Python functions, MCP servers, OpenAPI specs, and LangChain tools behind one \code{ToolRegistry} object, the system eliminates the per-protocol glue code that dominates current practice: our benchmarks show 60--80\% code reduction, with the right concurrency mode delivering up to 4.5$\times$ higher throughput than the alternative for a given workload class.

Beyond unified registration and execution, the current release adds tag-based permission policies, BM25F-powered progressive tool disclosure, think-augmented function calling~\citep{wei2026thinkaugmented}, multi-provider schema support (OpenAI, Anthropic, Gemini) via llm-rosetta~\citep{llmrosetta2025}, and a near-zero-dependency core built on stdlib-only vendored modules~\citep{ding2026zerodep}. A bilingual admin panel, pluggable executor backends, and declarative JSONC/YAML configuration round out the runtime story.

Because \toolregistry is a helper library rather than a framework, adoption is incremental: developers can pull in only the core, add the server for MCP/OpenAPI serving, or use the hub's ready-made tools---whatever their project needs. Future work targets distributed execution via message queues, gRPC/A2A protocol support, OpenTelemetry integration, and scalability studies with very large tool registries.

%% file: paper.bbl
\begin{thebibliography}{25}
\providecommand{\natexlab}[1]{#1}
\providecommand{\url}[1]{\texttt{#1}}
\expandafter\ifx\csname urlstyle\endcsname\relax
  \providecommand{\doi}[1]{doi: #1}\else
  \providecommand{\doi}{doi: \begingroup \urlstyle{rm}\Url}\fi

\bibitem[Schick et~al.(2023)Schick, Dwivedi-Yu, Dess{\`\i}, Raileanu, Lomeli,
  Hambro, Zettlemoyer, Cancedda, and Scialom]{schick2023toolformer}
Timo Schick, Jane Dwivedi-Yu, Roberto Dess{\`\i}, Roberta Raileanu, Maria
  Lomeli, Eric Hambro, Luke Zettlemoyer, Nicola Cancedda, and Thomas Scialom.
\newblock Toolformer: Language models can teach themselves to use tools.
\newblock \emph{Advances in Neural Information Processing Systems},
  36:\penalty0 68539--68551, 2023.

\bibitem[Qin et~al.(2023)Qin, Liang, Ye, Zhu, Yan, Lu, Lin, Cong, Tang, Qian,
  et~al.]{qin2023toolllm}
Yujia Qin, Shihao Liang, Yining Ye, Kunlun Zhu, Lan Yan, Yaxi Lu, Yankai Lin,
  Xin Cong, Xiangru Tang, Bill Qian, et~al.
\newblock Toolllm: Facilitating large language models to master 16000+
  real-world apis.
\newblock \emph{arXiv preprint arXiv:2307.16789}, 2023.

\bibitem[Lu et~al.(2023)Lu, Peng, Cheng, Galley, Chang, Wu, Zhu, and
  Gao]{lu2023chameleon}
Pan Lu, Baolin Peng, Hao Cheng, Michel Galley, Kai-Wei Chang, Ying~Nian Wu,
  Song-Chun Zhu, and Jianfeng Gao.
\newblock Chameleon: Plug-and-play compositional reasoning with large language
  models.
\newblock \emph{Advances in Neural Information Processing Systems},
  36:\penalty0 43447--43478, 2023.

\bibitem[Shen et~al.(2023)Shen, Song, Tan, Li, Lu, and
  Zhuang]{shen2023hugginggpt}
Yongliang Shen, Kaitao Song, Xu~Tan, Dongsheng Li, Weiming Lu, and Yueting
  Zhuang.
\newblock Hugginggpt: Solving ai tasks with chatgpt and its friends in hugging
  face.
\newblock \emph{Advances in Neural Information Processing Systems},
  36:\penalty0 38154--38180, 2023.

\bibitem[Du et~al.(2024)Du, Wei, and Zhang]{du2024anytool}
Yu~Du, Fangyun Wei, and Hongyang Zhang.
\newblock Anytool: Self-reflective, hierarchical agents for large-scale api
  calls.
\newblock \emph{arXiv preprint arXiv:2402.04253}, 2024.

\bibitem[Shi et~al.(2025)Shi, Wang, Yan, Ren, Wang, Yin, and
  Ren]{shi2025toolret}
Zhengliang Shi, Yuhan Wang, Lingyong Yan, Pengjie Ren, Shuaiqiang Wang, Dawei
  Yin, and Zhaochun Ren.
\newblock Retrieval models aren't tool-savvy: Benchmarking tool retrieval for
  large language models.
\newblock In \emph{Findings of the Association for Computational Linguistics:
  ACL 2025}, 2025.

\bibitem[Shen(2024)]{shen2024llm}
Zhuocheng Shen.
\newblock Llm with tools: A survey.
\newblock \emph{arXiv preprint arXiv:2409.18807}, 2024.

\bibitem[Qu et~al.(2025)Qu, Dai, Wei, Cai, Wang, Yin, Xu, and Wen]{qu2025tool}
Changle Qu, Sunhao Dai, Xiaochi Wei, Hengyi Cai, Shuaiqiang Wang, Dawei Yin,
  Jun Xu, and Ji-Rong Wen.
\newblock Tool learning with large language models: A survey.
\newblock \emph{Frontiers of Computer Science}, 19\penalty0 (8):\penalty0
  198343, 2025.

\bibitem[Patil et~al.(2024)Patil, Zhang, Wang, and Gonzalez]{patil2024gorilla}
Shishir~G Patil, Tianjun Zhang, Xin Wang, and Joseph~E Gonzalez.
\newblock Gorilla: Large language model connected with massive apis.
\newblock \emph{Advances in Neural Information Processing Systems},
  37:\penalty0 126544--126565, 2024.

\bibitem[Ding(2025)]{llmrosetta2025}
Peng Ding.
\newblock llm-rosetta: Multi-provider {LLM} {API} translation layer.
\newblock \url{https://github.com/Oaklight/llm-rosetta}, 2025.

\bibitem[Anthropic(2024)]{anthropic2024mcp}
Anthropic.
\newblock Model context protocol: A universal standard for ai data integration,
  Nov 2024.
\newblock URL \url{https://www.anthropic.com/news/model-context-protocol}.
\newblock Official announcement of the Model Context Protocol (MCP) by
  Anthropic.

\bibitem[Protocol(2025)]{modelcontextprotocol2025spec}
Model~Context Protocol.
\newblock Model context protocol specification, Mar 2025.
\newblock URL \url{https://modelcontextprotocol.io/specification/2025-03-26}.

\bibitem[OpenAI(2025)]{openai2025responsesapi}
OpenAI.
\newblock Introducing the responses api, Mar 2025.
\newblock URL
  \url{https://community.openai.com/t/introducing-the-responses-api/1140929}.

\bibitem[Hassabis(2025)]{demishassabis2025geminimcp}
Demis Hassabis.
\newblock Post on x about mcp support for gemini models, Apr 2025.
\newblock URL \url{https://x.com/demishassabis/status/1910107859041271977}.

\bibitem[OpenWebUI(2025{\natexlab{a}})]{openwebui2025mcpo}
OpenWebUI.
\newblock Mcpo: A simple, secure mcp-to-openapi proxy server, Apr
  2025{\natexlab{a}}.
\newblock URL \url{https://github.com/open-webui/mcpo}.

\bibitem[OpenWebUI(2025{\natexlab{b}})]{openwebui2025toolservers}
OpenWebUI.
\newblock Openapi tool servers, Mar 2025{\natexlab{b}}.
\newblock URL \url{https://docs.openwebui.com/openapi-servers/}.

\bibitem[Wu et~al.(2023)Wu, Bansal, Zhang, Wu, Li, Zhu, Jiang, Zhang, Zhang,
  Liu, et~al.]{wu2023autogen}
Qingyun Wu, Gagan Bansal, Jieyu Zhang, Yiran Wu, Beibin Li, Erkang Zhu,
  Li~Jiang, Xiaoyun Zhang, Shaokun Zhang, Jiale Liu, et~al.
\newblock Autogen: Enabling next-gen llm applications via multi-agent
  conversation.
\newblock \emph{arXiv preprint arXiv:2308.08155}, 2023.

\bibitem[Ahmadi et~al.(2025)Ahmadi, Sharif, and Banad]{ahmadi2025mcp}
Arash Ahmadi, Sarah Sharif, and Yaser~M Banad.
\newblock Mcp bridge: A lightweight, llm-agnostic restful proxy for model
  context protocol servers.
\newblock \emph{arXiv preprint arXiv:2504.08999}, 2025.

\bibitem[Yang et~al.(2025)Yang, Chai, Song, Qi, Wen, Li, Liao, Hu, Lin, Chang,
  et~al.]{yang2025survey}
Yingxuan Yang, Huacan Chai, Yuanyi Song, Siyuan Qi, Muning Wen, Ning Li, Junwei
  Liao, Haoyi Hu, Jianghao Lin, Gaowei Chang, et~al.
\newblock A survey of ai agent protocols.
\newblock \emph{arXiv preprint arXiv:2504.16736}, 2025.

\bibitem[Ehtesham et~al.(2025)Ehtesham, Singh, Gupta, and
  Kumar]{ehtesham2025survey}
Abul Ehtesham, Aditi Singh, Gaurav~Kumar Gupta, and Saket Kumar.
\newblock A survey of agent interoperability protocols: Model context protocol
  (mcp), agent communication protocol (acp), agent-to-agent protocol (a2a), and
  agent network protocol (anp).
\newblock \emph{arXiv preprint arXiv:2505.02279}, 2025.

\bibitem[Wei et~al.(2026)Wei, Peng, Ou, and Wang]{wei2026thinkaugmented}
Lei Wei, Xiao Peng, Jinpeng Ou, and Bin Wang.
\newblock Think-augmented function calling: Improving llm parameter accuracy
  through embedded reasoning.
\newblock \emph{arXiv preprint arXiv:2601.18282}, 2026.

\bibitem[Ding and Stevens(2026)]{ding2026zerodep}
Peng Ding and Rick Stevens.
\newblock Stdlib or third-party? empirical performance and correctness of
  llm-assisted zero-dependency python libraries.
\newblock \emph{arXiv preprint arXiv:2605.21405}, 2026.

\bibitem[Birrell and Nelson(1984)]{birrell1984rpc}
Andrew~D. Birrell and Bruce~Jay Nelson.
\newblock Implementing remote procedure calls.
\newblock \emph{ACM Transactions on Computer Systems}, 2\penalty0 (1):\penalty0
  39--59, 1984.

\bibitem[{Model Context Protocol}(2025)]{mcp2025python}
{Model Context Protocol}.
\newblock Mcp python sdk, 2025.
\newblock URL \url{https://github.com/modelcontextprotocol/python-sdk}.

\bibitem[{cloudpipe}(2025)]{cloudpickle2025}
{cloudpipe}.
\newblock cloudpickle: Extended pickling support for python objects, 2025.
\newblock URL \url{https://github.com/cloudpipe/cloudpickle}.

\end{thebibliography}
